\documentclass[11pt,a4paper,oneside]{article}
\usepackage{lineno,hyperref}

\usepackage{graphicx}              
\usepackage{epsfig}
\usepackage{amsmath}               
\usepackage{amssymb}
\usepackage{epstopdf}
\usepackage{verbatim}
\usepackage{mathptmx}
\usepackage{mathalpha}
\usepackage{mathtools}
\usepackage[scale=0.80]{geometry}
\usepackage{graphicx, times}
\usepackage{amsfonts}              
\usepackage{amsthm}                
\usepackage{multicol}
\usepackage{algorithm}
\usepackage{algorithmic}
\usepackage{varwidth}
\usepackage{parskip}
\usepackage{hyperref}
\usepackage{rotating}
\usepackage[numbers,sort&compress]{natbib}
\usepackage{multirow}
\usepackage{pdflscape}
\usepackage[numbers]{natbib}
\usepackage{caption}
\usepackage{xcolor}
\usepackage{color}
\usepackage{comment}
\usepackage{subcaption}
\newcommand*{\rom}[1]{\expandafter\@\romannumeral #1}

\newcommand{\bea}{\begin{eqnarray}}
	\newcommand{\eea}{\end{eqnarray}}
\newcommand{\bee}{\begin{eqnarray*}}
	\newcommand{\eee}{\end{eqnarray*}}

\begin{document}
\author{Khomesh R. Patle$^{}$\footnote{khomeshpatle5@gmail.com}, G. P. Singh$^{}$\footnote{gpsingh@mth.vnit.ac.in}, Romanshu Garg$^{}$\footnote{romanshugarg18@gmail.com}
\vspace{.3cm}\\
${}^{}$ Department of Mathematics,\\ Visvesvaraya National Institute of Technology, Nagpur, 440010, Maharashtra, India.
\vspace{.3cm}
\date{}}
	
\title{Dynamical constraints on variable vacuum energy in Brans-Dicke theory}
\maketitle
\begin{abstract} \noindent
In this research work, we investigate the late-time accelerated expansion of the universe within the framework of Brans-Dicke theory by considering dynamical vacuum energy models with a time-varying cosmological constant. Two vacuum energy models are studied, namely the hybrid vacuum law $\Lambda(t)=\alpha H^{2}+\beta\dot{H}$ and the power vacuum law $\Lambda(H)=\alpha_{1}H^{n}$, where $\alpha$, $\beta$, $\alpha_{1}$ and $n$ are free parameters. We derive analytical solutions for the Hubble parameter and other relevant cosmological quantities. The evolution of the deceleration parameter, the effective equation of state parameter, the cosmographic parameters, the behaviour of Om($\mathit{z}$) diagnostics and the present age of the universe are examined. Furthermore, the analysis of the $\omega_{\rm eff}-\omega'_{\rm eff}$ plane shows that the model evolves in the freezing region and the thermodynamic analysis confirms that the generalized second law of thermodynamics is satisfied within the power vacuum law model.
\end{abstract}

\section{Introduction}\label{sec:1}
Astrophysical observations across a wide range of redshifts~\cite{1998AJ....116.1009R,1999ApJ...517..565P,aghanim2020planck,de2000flat} provide compelling evidence that the universe is spatially flat and currently undergoing accelerated expansion. This late-time acceleration is attributed to an unknown component, commonly referred to as dark energy which constitutes approximately $68\%$ of the total cosmic energy density and is characterized by a negative pressure responsible for the observed acceleration. Despite extensive investigations, the fundamental nature of dark energy remains one of the most profound open problems in modern cosmology. The simplest type of dark energy is described by the constant vacuum energy density which is visualized as the cosmological constant~\cite{weinberg1989cosmological}. 
The most widely accepted model of the universe is the $\Lambda$ cold dark matter ($\Lambda$CDM) model.
Nevertheless, $\Lambda$CDM suffers from theoretical shortcomings, most notably the fine-tuning problem and the cosmic coincidence problem~\cite{carroll2001cosmological,di2021realm}.
In view of these issues, a variety of alternative cosmological models have been proposed in the literature. Among them, models featuring a time-varying vacuum energy density have attracted considerable attention as they offer a promising framework for describing both the early and late-time evolution of the universe within a unified approach.
Concurrently with the advancement of knowledge, a variety of modified gravity theories have attracted considerable attention and extensive investigation~\cite{harko2011f,nojiri2011unified,jimenez2017coincident,harko2010f,capozziello2011cosmography,bamba2010finite,singh505abc,mandal2023cosmic,patle2025accelerated,varela2025cosmological,kotambkar2017anisotropic,singh2025observational}, reflecting the dynamic nature of theoretical development in this field. 
\vspace{0.2cm}\\
The Brans-Dicke (BD) theory, initially proposed by Brans and Dicke~\cite{brans1961mach}, is a well-known scalar-tensor theory of gravity inspired by Mach’s principle and Dirac’s large number hypothesis. Unlike general relativity (GR), where gravitation is described solely by the curvature of spacetime, BD theory introduces an additional scalar field that dynamically governs the gravitational coupling alongside the metric tensor. In its original framework, the scalar field is massless and long-ranged which imposes stringent observational constraints on the BD coupling parameter $\omega_{BD}$.
One commonly adopted approach involves introducing a scalar-field potential which generates an effective mass for the scalar field and mimics the role of a cosmological constant. Another viable generalization of the theory incorporates an explicit cosmological constant that couples to the scalar field in the same manner as the Ricci curvature scalar in the original BD action. This formulation maintains the long-range nature of the scalar field and remains faithful to the foundational principles of BD gravity. The presence of the cosmological constant introduces additional dynamical freedom, leading to richer cosmological behaviours and new solutions that do not appear in the original BD framework. As a result, this scenario provides a promising theoretical setting for investigating late-time cosmic acceleration.
Scalar-tensor theories and in particular BD theory offer a compelling alternative to GR. In BD theory, the gravitational interaction is mediated by both a scalar field $\phi$ and the metric tensor $g_{\mu \nu}$ reflecting the influence of spacetime curvature together with the dynamics of the scalar field. Although the original formulation of BD theory does not include a cosmological constant, in this work we consider BD theory with a cosmological constant. A natural generalization of GR with a cosmological constant to the BD framework is obtained by replacing the Newtonian gravitational constant $\mathit{G}$ with a dynamical scalar field $\phi$ and introducing a kinetic term controlled by an arbitrary coupling parameter $\omega_{BD}$ known as the BD parameter.
\vspace{0.2cm}\\
The BD theory is characterized by a scalar field $\phi$, the spacetime metric $g_{\mu\nu}$ and the dimensionless coupling parameter $\omega_{BD}$. In this study, we treat the cosmological constant or equivalently the vacuum energy density $\rho_{\Lambda}$ as an intrinsic part of the theoretical framework. 
Allowing the vacuum energy to evolve dynamically provides a phenomenologically attractive generalization of the cosmological constant. Models based on time-dependent vacuum energy densities may capture richer cosmic dynamics, although their largely phenomenological nature and limited connection to fundamental theory make systematic theoretical and observational testing essential. This class of theories has regained significant attention due to its close connection with superstring-inspired models~\cite{uehara1982brans,pimentel1985exact,amendola2003cosmic,fabris2005late,banerjee2007holographic,sheykhi2010interacting,bhardwaj2023bianchi,singh2011bianchi,singh2013dynamic,singh2017bulk,ghaffari2018tsallis,maurya2020reply,samanta2025tachyonic,layden2024invariant}.
\vspace{0.2cm}\\
Bertolami made early attempts to model dark energy through a dynamical cosmological constant~\cite{bertolami1986time} and later by Ozer and Taha~\cite{ozer1987model}, who proposed that $\Lambda$ may vary with cosmic time. This idea has attracted significant interest, leading to a wide range of studies on cosmological models with a time-evolving $\Lambda$ term \cite{peebles1988cosmology,carvalho1992cosmological,bertolami2000nonminimal,barrow2006cosmologies}. Within the framework of quantum field theory, renormalization naturally introduces a running vacuum energy density often expressed as $\Lambda \varpropto H^{2}$ \cite{shapiro2000scaling}. A more realistic and observationally viable decay prescription was subsequently introduced by Wang and Meng \cite{wang2004can}. Despite the lack of a fundamental theory specifying the exact behaviour of vacuum energy, phenomenological models of decaying vacuum energy have proven to be effective tools in cosmological studies. Several works \cite{elizalde2005dark,borges2005friedmann,carneiro2008observational,costa2010cosmological,grande2011hubble,szydlowski2015cosmology,rezaei2019can} have investigated such models by considering $\Lambda(t)$ as a function of the scale factor or the Hubble parameter. These approaches indicate that decaying vacuum energy scenarios can successfully describe the accelerated expansion of the universe and may offer possible resolutions to both the cosmological constant problem and the coincidence problem. Although Einstein’s general theory of relativity has been remarkably successful, increasing attention has been devoted over the past two decades to its alternative formulations. 
\vspace{0.2cm}\\
For further insights into running vacuum models and recent theoretical advancements~\cite{moreno2020running,sola2022cosmological,moreno2022equation,mansoulie2025energy}. Early investigations by Overduin and Cooperstock~\cite{overduin1998evolution} examined the evolution of the scale factor in cosmological models featuring a variable vacuum energy density, considering several phenomenological forms such as $\Lambda = A t^{-l}$, $\Lambda = B a^{-m}$, $\Lambda = C t^{n}$ and $\Lambda = D q^{r}$, where $A$, $B$, $C$, $D$, $l$, $m$, $n$ and $r$ are constants. More recently, Rezaei et al.~\cite{rezaei2019can} adopted a phenomenological approach to parameterize the time dependence of the cosmological term by expressing $\Lambda(t)$ as a power-series expansion in terms of the Hubble parameter and its time derivatives, including forms such as $\Lambda(t) \varpropto H$, $\Lambda(t) \varpropto \dot{H}$ and $\Lambda(t) \varpropto H^{2}$. Furthermore, Singh and Solà~\cite{singh2021friedmann} studied Friedmann cosmology within the BD framework by considering a decaying vacuum energy scenario, where the vacuum energy density evolves according to the law $\Lambda(t) = \lambda + \sigma H$ with $\lambda$ and $\sigma$ being constants and $H$ denoting the Hubble parameter.
In recent years, Brans-Dicke theory with a cosmological constant has been widely studied as a promising framework for explaining dark energy phenomena~\cite{chand2016frw,tripathy2023cosmological,sen2001late,Karimkhani:2019acn,singh2020some,aditya2019observational,pasqua2019look}.
Motivated by the above studies, this work further explores dynamical vacuum energy scenarios within the framework of BD cosmology by adopting more general functional forms of the cosmological term. In particular, we examine two phenomenologically well-motivated models: a hybrid vacuum law~\cite{arbab1999frw}, $\Lambda(t)=\alpha H^{2}+\beta\dot{H}$ and a power vacuum law~\cite{overduin1998evolution}, $\Lambda(H)=\alpha_{1}H^{n}$, where $\alpha$, $\beta$, $\alpha_{1}$ and $n$ are treated as free parameters. The cosmological consequences of these models are analyzed within the BD framework, focusing on their implications for the background evolution of the universe.
\vspace{0.2cm}\\
This paper is organized into six sections. In Section (\ref{sec:2}), we present the field equations of BD theory for a homogeneous and isotropic FLRW spacetime and obtain a single evolution equation governing the Hubble parameter. Section (\ref{sec:3}) is devoted to the study of cosmological dynamics in the presence of a variable cosmological constant $\Lambda$, where both the hybrid vacuum law $\Lambda(t)$ model and the power vacuum law $\Lambda(H)$ model are examined. The main results and their phenomenological implications are discussed in Section (\ref{sec:4}), including the analysis of the deceleration parameter, the effective equation of state (EoS) parameter, the cosmographic parameters, the behaviour of Om($\mathit{z}$) diagnostics and evaluate the age of the universe. In this section (\ref{sec:5}), the trajectories of the $\omega_{\rm eff}-\omega'_{\rm eff}$ analysis are plotted for different values of the model parameter $n$ and the thermodynamic analysis of the model is also presented. Finally, the conclusions are summarized in Section (\ref{sec:6}).
\section{The field equations in BD theory}\label{sec:2}
Assuming a homogeneous and isotropic universe, the spacetime geometry for the flat case is described by the FLRW metric:
\begin{equation}{\label{1}}
	ds^{2}=-dt^{2}+a^{2}(t) \left[dr^{2}+r^{2} \left( d\theta^{2}+ sin^{2}\theta d\phi^{2}\right)\right],
\end{equation}
where the variables $(r, \theta, \phi)$ denote the comoving coordinates, $t$ is the proper cosmic time and $a(t)$ denotes the scale factor of the universe. We employ natural units such that $c=1$.  
\vspace{.2cm}\\
Accordingly, the BD action in the Jordan frame, where the matter Lagrangian $L_{m}$ does not couple to the scalar field, takes the form~\cite{uehara1982brans,kim2005brans,ozer2021gravitational}:
\begin{equation}{\label{2}}
	S=\int \left[\frac{1}{16\pi} \left( \phi R - \frac{\omega_{BD}}{\phi} \nabla_{\alpha} \phi \nabla^{\alpha} \phi  \right)-\rho_{\Lambda}+ L_{m} \right] \sqrt{-g} d^{4}x,
\end{equation}
In this framework, $\phi$ acts as an additional degree of freedom, functioning as the BD scalar field that is non-minimally coupled to the curvature scalar R. Since $\phi$ $\varpropto$ $G^{-1}$, its evolution reflects the variation of the gravitational coupling. The quantity $\omega_{BD}$ is the dimensionless BD parameter and $L_{m}$ denotes the matter Lagrangian density. The vacuum energy density corresponding to the cosmological constant is $\rho_{\Lambda} = \Lambda G^{-1} / 8\pi$ = $\Lambda\phi$$ /8\pi$. The original BD theory is recovered when $\Lambda=0$~\cite{brans1961mach}. Moreover, when the scalar field $\phi$ becomes constant, the theory reduces to GR with a cosmological constant. All remaining symbols have their conventional interpretation. 
\vspace{0.2cm}\\
Applying variational principles to the action (\ref{2}) with respect to the metric $g_{\mu \nu}$ and the scalar field $\phi$ results in the following set of field equations as
\begin{equation}{\label{3}}
	G_{\mu \nu} = R_{\mu \nu}-\frac{R}{2}g_{\mu \nu} = \frac{8\pi}{\phi}T_{\mu \nu}^{BD}+\frac{8\pi}{\phi}\tilde{T}_{\mu \nu}
\end{equation}
and
\begin{equation}{\label{4}}
	\nabla_{\alpha} \nabla^{\alpha} \phi= \frac{-8\pi}{(3+2\omega_{BD})}(4\rho_{\Lambda}- T_{\lambda}^{m\lambda}).
\end{equation}
The quantity $\tilde{T}_{\mu \nu}$ represents the total energy-momentum tensor, composed of the matter part and the vacuum energy density and is written as $T_{\mu \nu}^{m}-g_{\mu \nu}\rho_{\Lambda}$. Assuming a perfect fluid description, we write
\begin{equation}{\label{5}}
 \tilde{T}_{\mu \nu} = (\rho +p)u_{\mu}u_{\nu}+p g_{\mu \nu},
\end{equation}
with total energy density and pressure given by $\rho = \rho_{m}+\rho_{\Lambda}$ and $p = p_{m}+p_{\Lambda}$. Here, $\rho_{m}$ and $\rho_{\Lambda}$ denote the matter and vacuum energy densities, while $p_{m}$ and $p_{\Lambda}$ correspond to their respective pressures. For our analysis, we take the matter component to be dust (including dark matter), meaning $p_{m}=0$.
\vspace{0.2cm}\\
The BD scalar field contributes an energy-momentum tensor $T_{\mu \nu}^{BD}$, which is defined in Eq. (\ref{3}) as
\begin{equation}{\label{6}}
	T_{\mu \nu}^{BD} = \frac{1}{8\pi}\left[ \left(\nabla_{\mu}\nabla_{\nu}\phi-g_{\mu\nu}\nabla_{\alpha}\nabla^{\alpha}\phi \right)+ \frac{\omega_{BD}}{\phi} \left(\nabla_{\mu}\phi\nabla_{\nu}\phi-\frac{1}{2}g_{\mu\nu}\nabla_{\alpha}\phi\nabla^{\alpha}\phi \right)     \right].
\end{equation}
Using the metric form provided in Eq. (\ref{1}), the BD field equations (\ref{3}) take the following form:

\begin{equation}{\label{7}}
3H^{2}+3H\frac{\dot{\phi}}{\phi}-\frac{\omega_{BD}}{2} \frac{\dot{\phi}^{2}}{\phi^{2}} = \frac{8\pi}{\phi}\rho,
\end{equation}
\begin{equation}{\label{8}}
2\dot{H}+3H^{2}+2H\frac{\dot{\phi}}{\phi}+\frac{\ddot{\phi}}{\phi}+\frac{\omega_{BD}}{2} \frac{\dot{\phi}^{2}}{\phi^{2}} = -\frac{8\pi}{\phi}p,
\end{equation} 
\begin{equation}{\label{9}}  
\ddot{\phi}+3H\dot{\phi} = \frac{8\pi}{(3+2\omega_{BD})}(\rho-3p).
\end{equation}
In this expression, the overdot signifies differentiation with respect to the cosmic time `t', while the Hubble parameter is defined as $H=\dot{a}/a$. Using the Bianchi identity $\nabla_{\nu}G^{\mu\nu}=0$ in equation (\ref{3}), we obtain the following consistency relation.
\begin{equation}{\label{10}}
	\nabla_{\nu}\left(R^{\mu \nu}-\frac{R}{2}g^{\mu \nu}\right) =0= \nabla_{\nu} \left(\frac{8\pi}{\phi}T_{BD}^{\mu \nu}+\frac{8\pi}{\phi}\tilde{T}^{\mu \nu}\right).
\end{equation}
By imposing that the energy-momentum tensor $\tilde{T}_{\mu \nu}$ satisfies the standard conservation law $\tilde{T}^{\mu \nu}_{;\nu}=0$, we obtain the resulting expression  
\begin{equation}{\label{11}}
\dot{\rho}_{m}+3\frac{\dot{a}}{a}(\rho_{m}+p_{m})= -\dot{\rho}_{\Lambda}.
\end{equation}
We consider the energy-momentum tensor of the BD scalar field, $T_{\mu \nu}^{BD}$ to behave like a perfect fluid. Accordingly, it is expressed as $T^{BD}_{\mu \nu} = (\rho_{BD} +p_{BD})u_{\mu}u_{\nu}+p_{BD} g_{\mu \nu}$, where

\begin{equation}{\label{12}}
\rho_{BD}=\frac{1}{8\pi}\left(\frac{\omega_{BD}}{2} \frac{\dot{\phi}^{2}}{\phi}-3H\dot{\phi}\right),
\end{equation}
\begin{equation}{\label{13}}
p_{BD}=\frac{1}{8\pi}\left(\frac{\omega_{BD}}{2} \frac{\dot{\phi}^{2}}{\phi}+2H\dot{\phi}+\ddot{\phi}\right).
\end{equation}
Using equation (\ref{11}), the Bianchi identity in Eq. (\ref{10}) leads to
\begin{equation}{\label{14}}
\nabla_{\nu} \left(\frac{8\pi}{\phi}T_{BD}^{\mu \nu}\right)+8\pi \tilde{T}^{\mu \nu} \nabla_{\nu}\left(\frac{1}{\phi}\right)=0,
\end{equation}
which can be simplified to
\begin{equation}{\label{15}}
\dot{\rho}_{BD}+3H(\rho_{BD}+p_{BD})=\left(\frac{\dot{\phi}}{\phi}\right) (\rho_{BD}+\rho_{m}+\rho_{\Lambda}).
\end{equation}
As described in Refs.~\cite{pimentel1985exact,banerjee2007holographic,sheykhi2010interacting,johri1994cosmological}, we adopt a power-law evolution for the BD scalar field, taking $\phi=\phi_{0}a(t)^\epsilon$, where $\phi_{0}$ and $\epsilon$ are constants. The parameter $\epsilon$ is chosen to be small so that variations in $\mathit{G}$ remain observationally acceptable. For sufficiently large $\omega_{BD}$, the combination $\epsilon$$\omega_{BD}$ naturally stays of order unity~\cite{banerjee2007holographic}, allowing the scalar field to influence cosmological dynamics. In addition, the Cassini experiment imposes a strong lower bound on $\omega_{BD}$.
Interestingly, a similar power-law form for the BD scalar was employed in Ref.~\cite{peracaula2018brans} to investigate BD cosmology in the presence of a cosmological term. The study demonstrated that such a choice significantly improves the agreement with cosmological observations~\cite{perez2018brans}. This suggests that adopting the same assumption could be equally advantageous when analyzing different time-varying vacuum energy density scenarios in BD theory.
\vspace{0.2cm}\\
Accordingly, when the power-law expression for the BD scalar is applied, equation (\ref{7}) takes the form
\begin{equation}{\label{16}}
H^{2}=\frac{2}{(6+6\epsilon-\omega_{BD}\epsilon^{2})}\left(\frac{8\pi}{\phi}\right)(\rho_{m}+\rho_{\Lambda}).
\end{equation}
As indicated by equation (\ref{16}), the general relativistic cosmology is retrieved in the limit $\epsilon \rightarrow 0$. Under the dust condition $p_{m}=0$, equations (\ref{11}) and (\ref{16}) can be combined to produce a single evolution equation governing the Hubble parameter as
\begin{equation}{\label{17}}
	\dot{H}+\frac{(3+\epsilon)}{2}H^{2}=\frac{3}{(6+6\epsilon-\omega_{BD}\epsilon^{2})}\left(\frac{8\pi}{\phi}\right)\rho_{\Lambda}= \frac{3\Lambda}{(6+6\epsilon-\omega_{BD}\epsilon^{2})},
\end{equation}
here, $H=\dot{a}/a$ represents the Hubble parameter, while $\rho_{\Lambda} = \Lambda \mathit{G}^{-1} / 8\pi$ = $\Lambda\phi$$ /8\pi$. The equation can be integrated once the explicit form of $\Lambda$ is provided. In the next section, we derive the solution of equations (\ref{17}) by adopting a time-varying cosmological constant.
\section{Dynamics in the presence of a variable cosmological constant ($\Lambda$)}\label{sec:3}
In this work, we investigate the cosmological implications of a time-varying vacuum term $\Lambda(t)$ within the framework of BD theory, where both the scalar field and the dynamical $\Lambda$ component contribute to the evolution of the universe. Motivated by dynamical vacuum energy scenarios and renormalization-group arguments in curved spacetime, we first consider the hybrid vacuum law defined by $\Lambda(t)=\alpha H^{2}+\beta\dot{H}$, which incorporates both the expansion rate and its time variation, where $\alpha$ and $\beta$ are constants. Hybrid-type $\Lambda(t)$ models have been discussed in the literature on dynamical dark energy~\cite{arbab1999frw,al1998cosmological}, where the presence of both $H^{2}$ and $\dot{H}$ terms leads to modified cosmological dynamics through their coupling with the BD scalar field $\phi(t)$. 
As a second model, we adopt the power vacuum law, given by $\Lambda(H)=\alpha_{1}H^{n}$, where $\alpha_{1}$ and $n$ are constants. Power-law forms of vacuum energy have been widely investigated in the context of dynamical vacuum models~\cite{overduin1998evolution,myrzakulov2024signature}, as they provide a phenomenological description of a time-dependent vacuum component driven by the cosmic expansion. In light of these studies, we extend our analysis by examining both the hybrid $\Lambda(t)$ model and the power-law $\Lambda(H)$ model within the BD framework. These two models are treated as distinct submodels to explore their implications for cosmological evolution.
\subsection{Hybrid vacuum law $\Lambda(t)$ model}\label{sec:3.1}
We assume that the cosmological term is governed by a hybrid vacuum prescription, in which it depends on both the quadratic Hubble parameter and its time derivative (hereafter referred to as the $\Lambda_{HVL}$ model). This form of the time-varying cosmological term can be written ~\cite{arbab1999frw} as
\begin{equation}{\label{18}}
 \Lambda(t)=\alpha H^{2}+\beta\dot{H},
\end{equation} 
where $\alpha$ and $\beta$ are dimensionless constants. Substituting equation (\ref{18}) into equation (\ref{17}), we obtain the modified cosmological field equations governing the dynamics of the model:
\begin{equation}{\label{19}}
\left(1-\frac{3\beta}{(6+6\epsilon-\omega_{BD}\epsilon^{2})}\right)\dot{H}+\left(\frac{(3+\epsilon)}{2}-\frac{3\alpha}{(6+6\epsilon-\omega_{BD}\epsilon^{2})}\right)H^{2}=0,
\end{equation} 
here the dot denotes $dH/dt$. Solving equation (\ref{19}), we find
\begin{equation}{\label{20}}
H(z)=H_{0}(1+z)^{\left(\frac{(3+\epsilon)(6+6\epsilon-\omega_{BD}\epsilon^{2})-6\alpha}{2(6+6\epsilon-\omega_{BD}\epsilon^{2}-3\beta)}\right)}.
\end{equation} 
In this expression, $H_{0}$ represents the present value of the Hubble parameter at $t=0$, while the redshift is defined by $z=(a_{0}/a)-1$ with $a_{0}=1$ assumed henceforth. Employing $H=\dot{a}/a$, the scale factor can be expressed as 
\begin{equation}{\label{21}}
a=\left[ \left(\frac{(3+\epsilon)(6+6\epsilon-\omega_{BD}\epsilon^{2})-6\alpha}{2(6+6\epsilon-\omega_{BD}\epsilon^{2}-3\beta)}\right) H_{0}t \right]^{\left(\frac{(3+\epsilon)(6+6\epsilon-\omega_{BD}\epsilon^{2})-6\alpha}{2(6+6\epsilon-\omega_{BD}\epsilon^{2}-3\beta)}\right)}.
\end{equation} 
\vspace{0.1cm}\\
We obtain a power-law solution for the scale factor, indicating that the cosmic evolution exhibits deceleration, marginal inflation or acceleration depending on whether $\left(\frac{(3+\epsilon)(6+6\epsilon-\omega_{BD}\epsilon^{2})-6\alpha}{2(6+6\epsilon-\omega_{BD}\epsilon^{2}-3\beta)}\right) > 1$, $\left(\frac{(3+\epsilon)(6+6\epsilon-\omega_{BD}\epsilon^{2})-6\alpha}{2(6+6\epsilon-\omega_{BD}\epsilon^{2}-3\beta)}\right)=1$ or $\left(\frac{(3+\epsilon)(6+6\epsilon-\omega_{BD}\epsilon^{2})-6\alpha}{2(6+6\epsilon-\omega_{BD}\epsilon^{2}-3\beta)}\right) < 1$ respectively. 
\vspace{.2cm}\\

\subsection{Power vacuum law $\Lambda(H)$ model}\label{sec:3.2}
We next consider a second class of dynamical vacuum model in which the cosmological term follows a power-law dependence on the Hubble parameter, hereafter referred to as the $\Lambda_{PVL}$ model. In this approach, the time-varying cosmological term is parameterized~\cite{overduin1998evolution,myrzakulov2024signature} as
\begin{equation}{\label{25}}
\Lambda(H)=\alpha_{1}H^{n},
\end{equation}
where $\alpha_{1}$ is a constant and $n$ denotes the power-law index. Other values of $n$ provide a broader phenomenological framework capable of describing different decay rates of the vacuum energy and their impact on cosmic evolution. 
Substituting equation (\ref{25}) into (\ref{17}), we obtain the corresponding modified field equations governing the dynamics of the power vacuum model:
\begin{equation}{\label{26}}
H^{\prime}+\left(\frac{3+\epsilon}{2}\right)\frac{H}{a}=\frac{3\alpha_{1}}{(6+6\epsilon-\omega_{BD}\epsilon^{2})}\frac{H^{(n-1)}}{a},
\end{equation}
where a prime represents $dH/da$. With the definition $\Lambda_{0}=3H_{0}^{2}\Omega_{\Lambda 0}$, equation (\ref{25}) leads to $\alpha_{1}=3H_{0}^{2-n}\Omega_{\Lambda 0}$, where the subscript `$0$' refers to the present epoch. Hence, Eq. (\ref{26}) takes the form
\begin{equation}{\label{27}}
H^{\prime}+\left(\frac{3+\epsilon}{2}\right)\frac{H}{a}=\frac{9H_{0}^{2-n}\Omega_{\Lambda 0}}{(6+6\epsilon-\omega_{BD}\epsilon^{2})}\frac{H^{(n-1)}}{a},
\end{equation}
Equation (\ref{27}) admits the following solution when written in terms of the redshift as
\begin{equation}{\label{28}}
H(z)=H_{0}\left[\frac{18\Omega_{\Lambda 0}}{(3+\epsilon)(6+6\epsilon-\omega_{BD}\epsilon^{2})}+\left(1-\frac{18\Omega_{\Lambda 0}}{(3+\epsilon)(6+6\epsilon-\omega_{BD}\epsilon^{2})}\right)(1+z)^{\gamma}\right]^\frac{1}{(2-n)},
\end{equation}
where $\gamma=\frac{(3+\epsilon)(2-n)}{2}$. \\
By introducing the dimensionless Hubble parameter $E(z)=H/H_{0}$. Therefore, equation (\ref{28}) can be cast into the form
\begin{equation}{\label{29}}
E^{(2-n)}(z)= \tilde{\Omega}_{\Lambda 0}+\tilde{\Omega}_{m 0}(1+z)^{\gamma},
\end{equation}
where
\begin{equation}{\label{30}}
\tilde{\Omega}_{\Lambda 0}=1-\tilde{\Omega}_{m 0}= \frac{18\Omega_{\Lambda 0}}{(3+\epsilon)(6+6\epsilon-\omega_{BD}\epsilon^{2})}.
\end{equation}
\vspace{.2cm}\\
It follows from the above equations that the conventional BD model with a constant cosmological term is recovered by setting $n=0$.
Furthermore, choosing $\epsilon=0$ leads to the standard $\Lambda$CDM regime, characterized by the usual scaling law of non-relativistic matter and a strictly constant vacuum energy density.
\vspace{0.1cm}\\
\section{Results and phenomenological discussion}\label{sec:4}
In this section, we investigate the cosmological implications of the proposed vacuum energy models by examining their expansion dynamics: 
\subsection{ Deceleration parameter }\label{sec:4.1}
\begin{center}
	\begin{figure}
		\includegraphics[width=17.5cm, height=8cm]{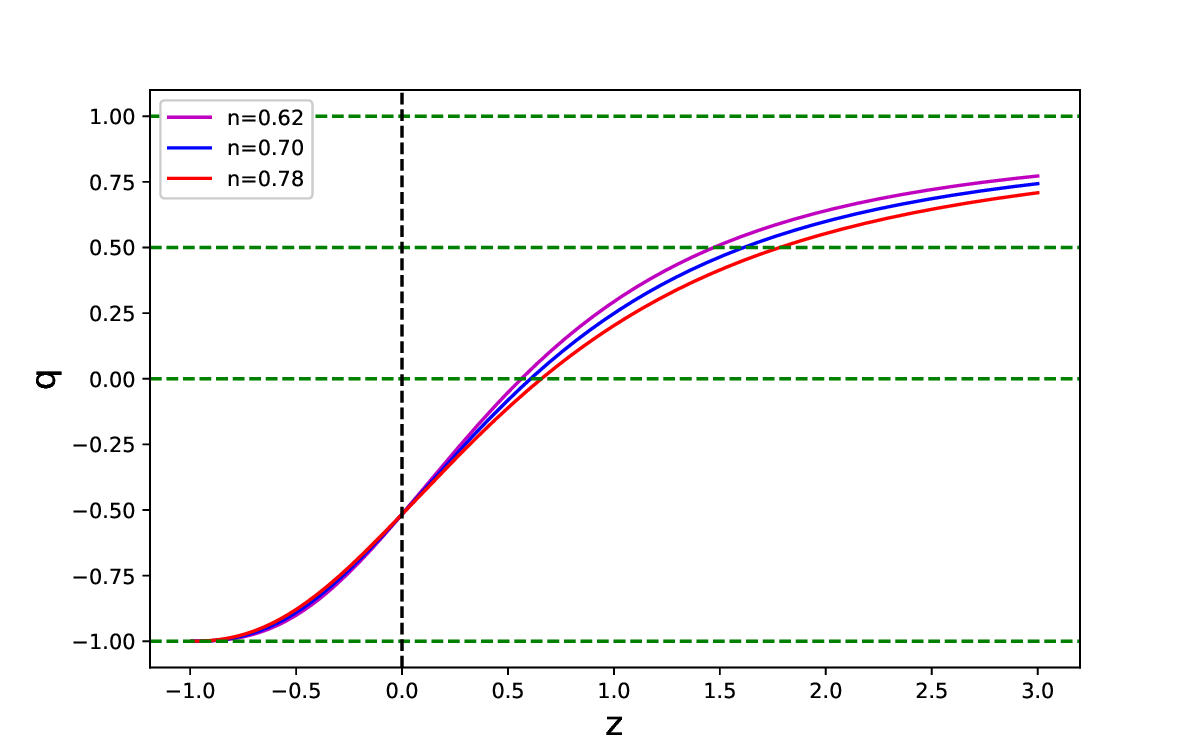}
		\caption{Deceleration parameter $q(z)$ versus $z$.} 
		\label{fig:1}
	\end{figure}
\end{center}

The deceleration parameter $(q)$ is one of the key quantities governing the expansion dynamics of the universe. It provides a direct measure of whether the cosmic expansion is accelerating or decelerating. Specifically, $ q<0 $ corresponds to an accelerated expansion phase, whereas $ q>0 $ indicates a decelerated expansion. Values of the deceleration parameter below $q=-1$ signify a super-accelerated expansion regime. Observationally relevant expansion histories of the universe are characterized by distinct values of $q$, namely $q=-1$, $q=\frac{1}{2}$ and $q=1$ which correspond to de Sitter, matter-dominated and radiation-dominated eras, respectively. 
In cosmological studies, the deceleration parameter $q$ plays a fundamental role in probing the transition from a decelerating to an accelerating universe. This dimensionless quantity is defined as
\begin{equation}{\label{22}}
	q=-1-\frac{\dot{H}}{H^{2}}.
\end{equation} 
\textbf{For $\Lambda_{HVL}$ model:} 
By inserting equation (\ref{20}) into equation (\ref{22}), we find that the deceleration parameter $q=-1+\left(\frac{(3+\epsilon)(6+6\epsilon-\omega_{BD}\epsilon^{2})-6\alpha}{2(6+6\epsilon-\omega_{BD}\epsilon^{2}-3\beta)}\right)$ is constant. This result indicates that the model predicts a purely decelerating $(q>0)$, marginally inflating $(q=0)$, or accelerating $(q<0)$ universe, depending on whether the expression in parentheses is greater than, equal to, or less than unity. Consequently, the model does not allow for the observed transition from an early decelerated epoch to the present accelerated phase and is therefore phenomenologically disfavored. 
In general, a time-dependent deceleration parameter is required to capture such a phase transition.
\vspace{0.2cm}\\
\textbf{For $\Lambda_{PVL}$ model:}
Using the equation (\ref{28}) into equation (\ref{22}), the deceleration parameter $q$ is given by
\begin{equation}{\label{31}}
	q(z)=-1+\frac{\gamma \tilde{\Omega}_{m 0} (1+z)^{\gamma}}{(2-n)\left[ \tilde{\Omega}_{\Lambda 0}+\tilde{\Omega}_{m 0}(1+z)^{\gamma}\right]}.
\end{equation}
The present-day value of the deceleration parameter, corresponding to $z=0$ is then obtained as
\begin{equation}{\label{32}}
	q_{0}=-1+\left(\frac{(3+\epsilon)(6+6\epsilon-\omega_{BD}\epsilon^{2})-18\Omega_{\Lambda 0}}{2(6+6\epsilon-\omega_{BD}\epsilon^{2})}\right).
\end{equation}
The corresponding transition redshift, $z_{tr}$ defined by the condition $q(z_{tr})=0$ is given by
\begin{equation}{\label{33}}
	z_{tr}=\left[\frac{2 \tilde{\Omega}_{\Lambda 0}}{(1+\epsilon)\tilde{\Omega}_{m 0}}\right]^{1/\gamma}-1.
\end{equation}
Figure $(\ref{fig:1})$ displays the evolution of the deceleration parameter $q(z)$ for the $\Lambda_{PVL}$ model with three values of the parameter $n$. The plot is obtained using appropriate model parameter values: $H_{0}=67.8$ Km/(sec.Mpc)~\cite{aghanim2020planck}, $\epsilon=0.816$, $\omega_{BD}=9.92$ and $\Omega_{\Lambda 0}=0.679$. The figure clearly shows a transition from an early decelerated expansion phase to the current accelerated phase of the universe. 
The corresponding transition redshift $z_{tr}$ and the present-day values of the deceleration parameter $q_{0}$ are summarized in Table~\ref{table:1}.
The negative value of $q_{0}$ confirms that the universe is presently undergoing accelerated expansion at $z=0$. 
The behaviour of $q$ shown in Fig.$(\ref{fig:1})$ indicates that the model exhibits a transition to an accelerated expansion regime at late times with $q<0$ and in the far future $(z \to -1)$ it asymptotically approaches a de Sitter phase characterized by $q=-1$. This overall evolution of the deceleration parameter provides strong confirmation that the model is well suited to describe the observed late-time acceleration of the universe.
\begin{table}[htbp]
	\centering
	\begin{tabular}{|c|c|c|c|c|c|c|c|c|c|}
		\hline
		Values of $n$ &  $q_{0}$ & $z_{tr}$ & $\omega_{0}$ & $j_{0}$ & $s_{0}$ & $t_{0}$(Gyr) \\
		\hline
		0.62 &  $-0.5162$ & $0.563$ & $-0.6775$ & $0.9676$ & $-0.6581$ & $12.67$ \\
		\hline
		0.70  &  $-0.5162$ & $0.612$ & $-0.6775$ & $0.9125$ & $-0.5530$ & $12.89$ \\
		\hline
		0.78  &  $-0.5162$ & $0.656$ & $-0.6775$ & $0.8573$ & $-0.4562$ & $13.12 $ \\
		\hline
	\end{tabular} 
	\caption{Present-day values of key cosmological parameters and the age of the universe for different values of the model parameter $n$.}
	\label{table:1}
\end{table}
\subsection{Effective EoS parameter}\label{sec:4.2}
\begin{center}
	\begin{figure}
		\includegraphics[width=17.5cm, height=8cm]{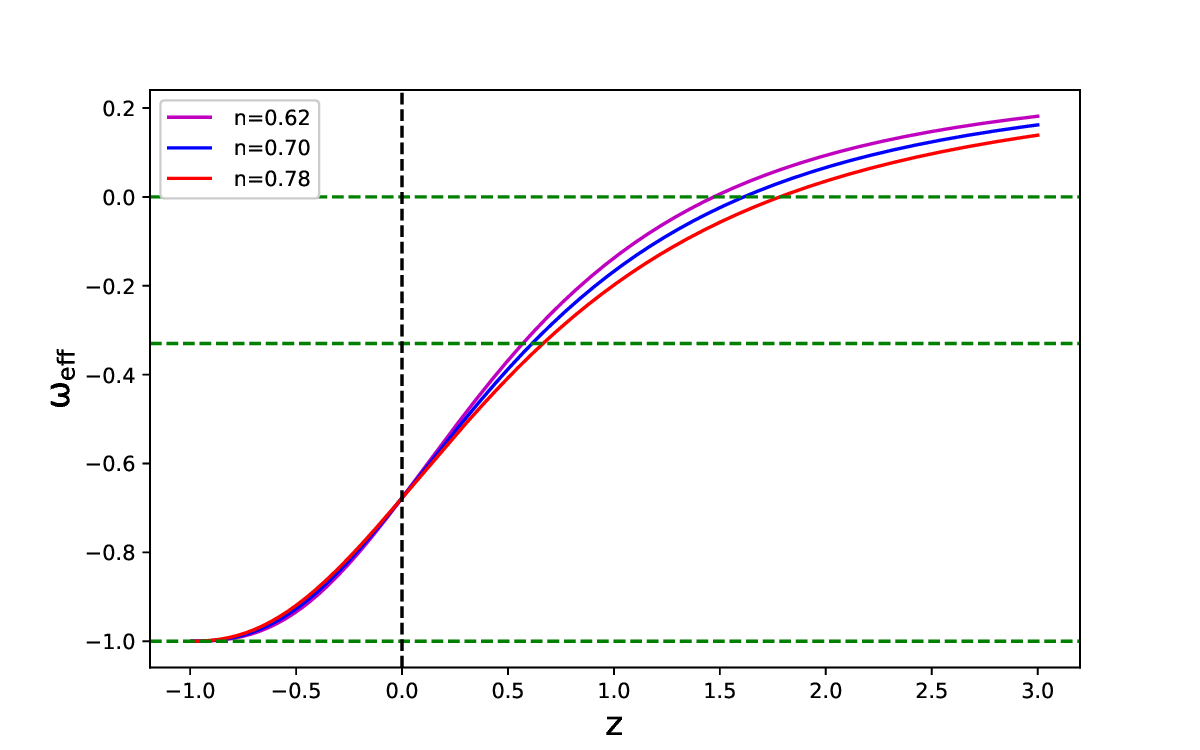}
		\caption{Effective EoS parameter ($\omega_{eff}$) versus $z$.} 
		\label{fig:2}
	\end{figure}
\end{center}

The effective equation of state parameter $(\omega_{eff})$ provides a unified description of the dynamical behavior of the universe across different evolutionary stages. In this framework, $\omega_{eff}=0$ corresponds to a matter-dominated epoch, $\omega_{eff}=\frac{1}{3}$ describes the radiation-dominated era and  $\omega_{eff}=-1$ characterizes vacuum energy associated with the de Sitter phase of the $\Lambda$CDM model. An accelerated expansion of the universe occurs when $\omega_{eff} < -\frac{1}{3}$. This accelerating regime includes the quintessence domain $(-1 < \omega_{eff} < -\frac{1}{3})$ and the phantom domain $(\omega_{eff} < -1 )$.
In what follows, we will refer to the relevant quantity as the effective equation of state (EoS) parameter $\omega_{eff}$, which is defined by
\begin{equation}{\label{34}}
	\omega_{eff}=-1-\frac{2}{3}\frac{a H^{\prime}}{H}.
\end{equation}
Within the framework of the $\Lambda_{PVL}$ model, the following expression is derived:
\begin{equation}{\label{35}}
	\omega_{eff}=-1+\frac{2\gamma \tilde{\Omega}_{m 0} (1+z)^{\gamma}}{(6-3n)\left[ \tilde{\Omega}_{\Lambda 0}+\tilde{\Omega}_{m 0}(1+z)^{\gamma}\right]}.
\end{equation}
At $z=0$, the above equation yields the present value of the effective EoS parameter as
\begin{equation}{\label{36}}
	\omega_{eff}(z=0)=-1+\frac{(3+\epsilon)(6+6\epsilon-\omega_{BD}\epsilon^{2})-18\Omega_{\Lambda 0}}{3(6+6\epsilon-\omega_{BD}\epsilon^{2})}.
\end{equation}
Figure~(\ref{fig:2}) depicts the redshift evolution of the effective dark energy EoS parameter $\omega_{eff}$ for three values of the parameter $n$. 
At the present epoch ($z=0$), the effective EoS parameter indicates a quintessence-like dark energy behaviour and the corresponding present-day values are provided in Table~\ref{table:1}. The evolution of $\omega_{eff}$ shows that the model undergoes an accelerated expansion phase at late times, characterized by $-1 < \omega_{eff} < -\frac{1}{3}$, while in the asymptotic future $(z \to -1)$ it smoothly approaches the $\Lambda$CDM limit with $\omega_{eff} \to -1$. This evolution supports the viability of the model in explaining the observed cosmic acceleration.
\subsection{Cosmographic parameters}\label{sec:4.3}
\begin{figure}[!htb]
	\captionsetup{skip=0.4\baselineskip,size=footnotesize}
	\begin{minipage}{0.50\textwidth}
		\centering
		\includegraphics[width=8.2cm,height=7cm]{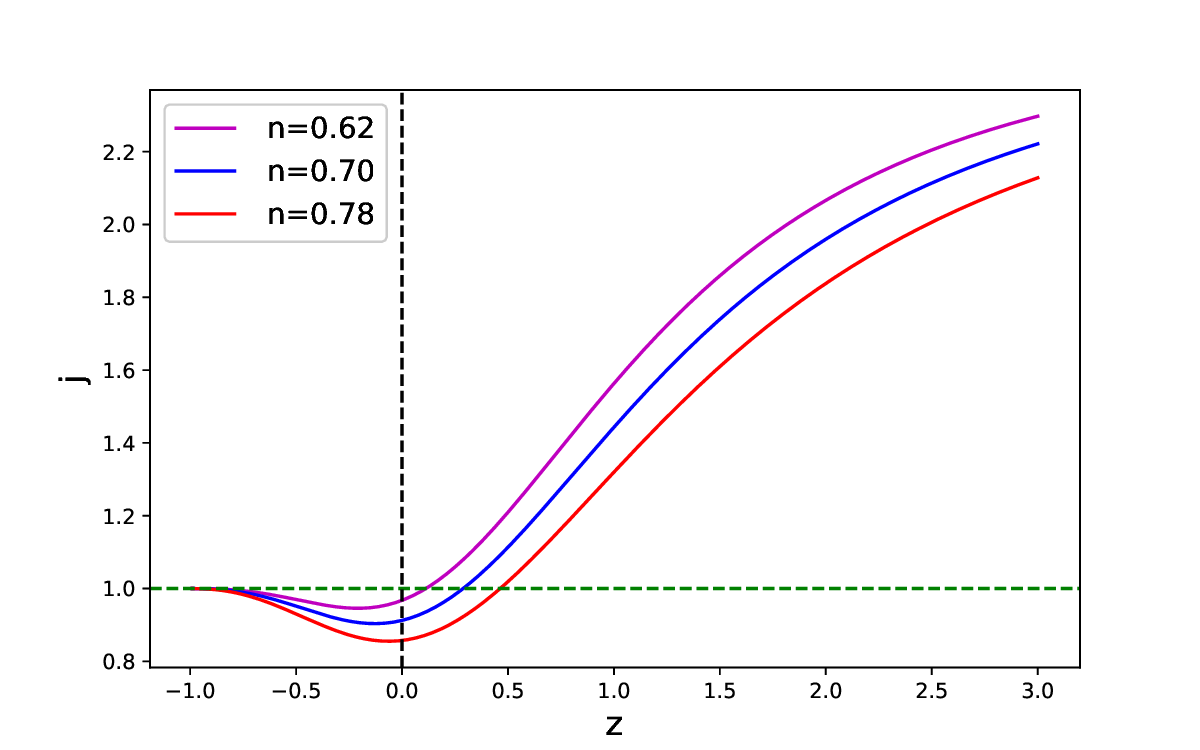}
		\caption{Jerk parameter ($\mathit{j}$) versus $\mathit{z}$.}
		\label{fig:3}
	\end{minipage}\hfill
	\begin{minipage}{0.50\textwidth}
		\centering
		\includegraphics[width=8.2cm,height=7cm]{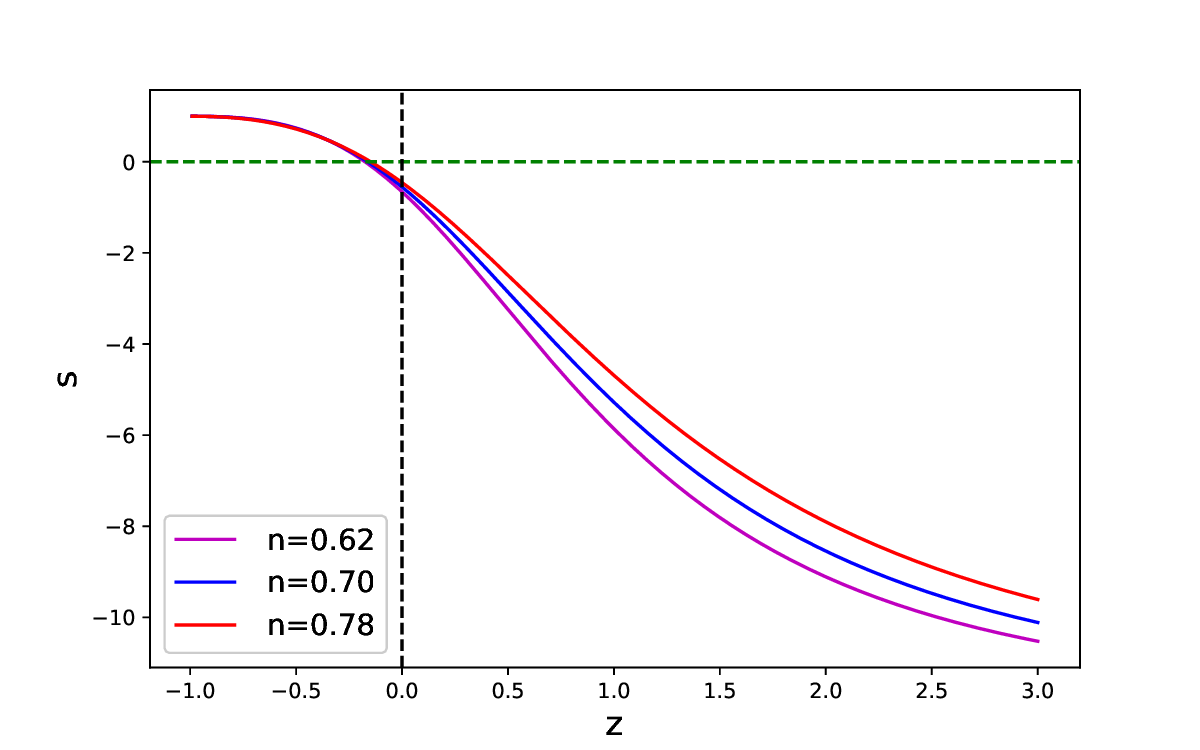}
		\caption{Snap parameter ($\mathit{s}$) versus $\mathit{z}$.}
		\label{fig:4}
	\end{minipage}
\end{figure}

The kinematic evolution of the universe can be effectively described using higher-order cosmographic parameters such as the jerk $\mathit{(j)}$ and snap $\mathit{(s)}$, which are defined in terms of the scale factor and its successive time derivatives. 
Cosmography was first introduced by Weinberg~\cite{weinberg2008cosmology} as a model-independent framework to investigate the expansion history of the universe by performing a Taylor expansion of the scale factor around the present epoch $t_{0}$.
\vspace{0.2cm}\\
Prior to the observational discovery of cosmic acceleration, the Hubble parameter $(H)$ was regarded as a time-dependent quantity reflecting the evolving expansion rate of the universe.
In this work, we extend the analysis to higher-order kinematic quantities by examining the jerk and snap parameters, also referred to as jolt and jounce, respectively. The study of these parameters provides deeper insight into the dynamical evolution of the universe and helps to distinguish between different cosmological scenarios~\cite{mukherjee2016parametric,visser2004jerk}:
\begin{equation}{\label{39}}
	\mathit{j}=\frac{1}{aH^{3}}\left(\frac{d^{3}a}{dt^{3}}\right),\ \ \mathit{s}=\frac{1}{aH^{4}}\left(\frac{d^{4}a}{dt^{4}}\right).
\end{equation}
\vspace{.1cm}\\
Using equation~(\ref{39}), we evaluate the jerk and snap parameters as functions of redshift~\cite{wang2009probing}.
\begin{equation}{\label{40}}
	j(z)=q(z)\left(2q(z)+1\right)+\frac{dq}{dz}(1+z), \  \ \  \   s(z)=-j(z)\left(3q(z)+2\right)-\frac{dj}{dz}(1+z).
\end{equation}
Figures~(\ref{fig:3}) and~(\ref{fig:4}) illustrate the redshift evolution of the jerk and snap parameters, respectively for the $\Lambda_{PVL}$ model with three values of the parameter $n$. The values of the jerk parameter $\mathit{j_{0}}$ and the snap parameter $\mathit{s_{0}}$ at the present epoch are provided in Table~\ref{table:1}. The deviation of $j_{0}$ from the $\Lambda$CDM value  $\mathit{j_{0}}= 1 $, indicates a departure from the standard cosmological model at the present epoch. Nevertheless, the overall evolution of these parameters suggests that the model gradually converges toward the $\Lambda$CDM limit at late times, highlighting its compatibility with standard cosmology in the asymptotic future.
\begin{center}
	\begin{figure}
		\includegraphics[width=15.5cm, height=7cm]{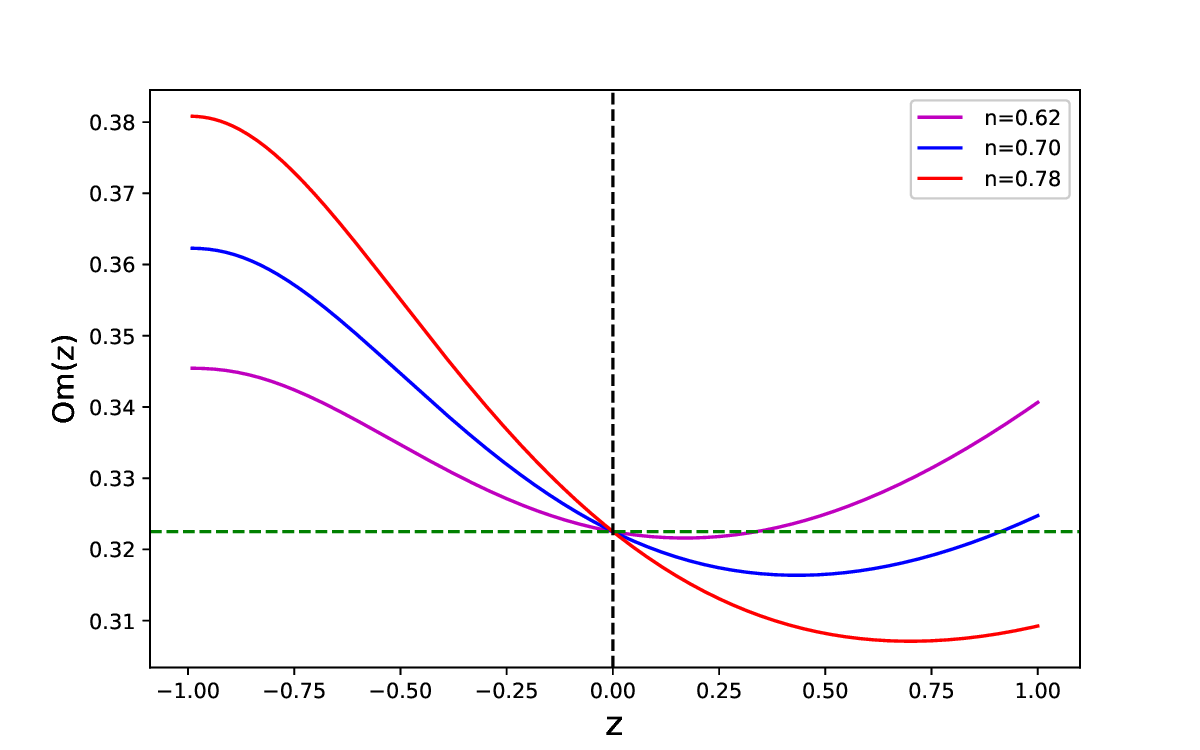}
		\caption{Om($\mathit{z}$) diagnostics versus $z$.} 
		\label{fig:5}
	\end{figure}
\end{center}
\subsection{Om($\mathit{z}$) diagnostics}\label{sec:4.4}
The Om($\mathit{z}$) diagnostics serves as a useful method for distinguishing among dark energy cosmological models~\cite{sahni2008two}. For flat universe, it is defined as
\begin{equation} {\label{41}}
	Om(z) = \frac{\frac{H^{2}(z)}{H^{2}_{0}}-1}{(1+z)^{3}-1}.
\end{equation}
The slope of Om($\mathit{z}$) provides insight into the nature of dark energy: a positive slope signifies phantom-like behaviour, whereas a negative slope corresponds to quintessence-like behaviour. The constant behaviour of Om($\mathit{z}$) characterizes the $\Lambda$CDM model. As shown in Fig.~(\ref{fig:5}), the evolution of Om($\mathit{z}$) diagnostic for the $\Lambda_{\mathrm{PVL}}$ model with three values of the parameter $n$ clearly indicates that the model exhibits quintessence-like behaviour at the present epoch ($z=0$).
\subsection{Age of the Universe}\label{sec:4.5}
The cosmic age `$t(z)$' as a function of redshift $ \mathit{z} $ can be expressed by the integral relation~\cite{tong2009cosmic}
\begin{equation} {\label{42}}
t(z) = \int_{z}^{\infty} \frac{d\tilde{z}}{(1+\tilde{z})H(\tilde{z})}.
\end{equation}
For the power vacuum law ($\Lambda_{\mathrm{PVL}}$) model, the reconstructed Hubble parameter $H(z)$ given in Eq.~(\ref{28}) is substituted into the above expression and evaluated at the present epoch ($z=0$) to estimate the age of the universe for three values of the parameter $n$. The resulting present cosmic age $t_{0}$ is summarized in Table~\ref{table:1}.
\section{Dynamical and thermodynamic analysis}\label{sec:5}
\subsection{The $\omega_{\rm eff}-\omega'_{\rm eff}$ analysis}\label{sec:5.1}
\begin{figure}[htbp]
	\centering
	
	\begin{subfigure}{0.32\textwidth}
		\centering
		\includegraphics[width=\linewidth]{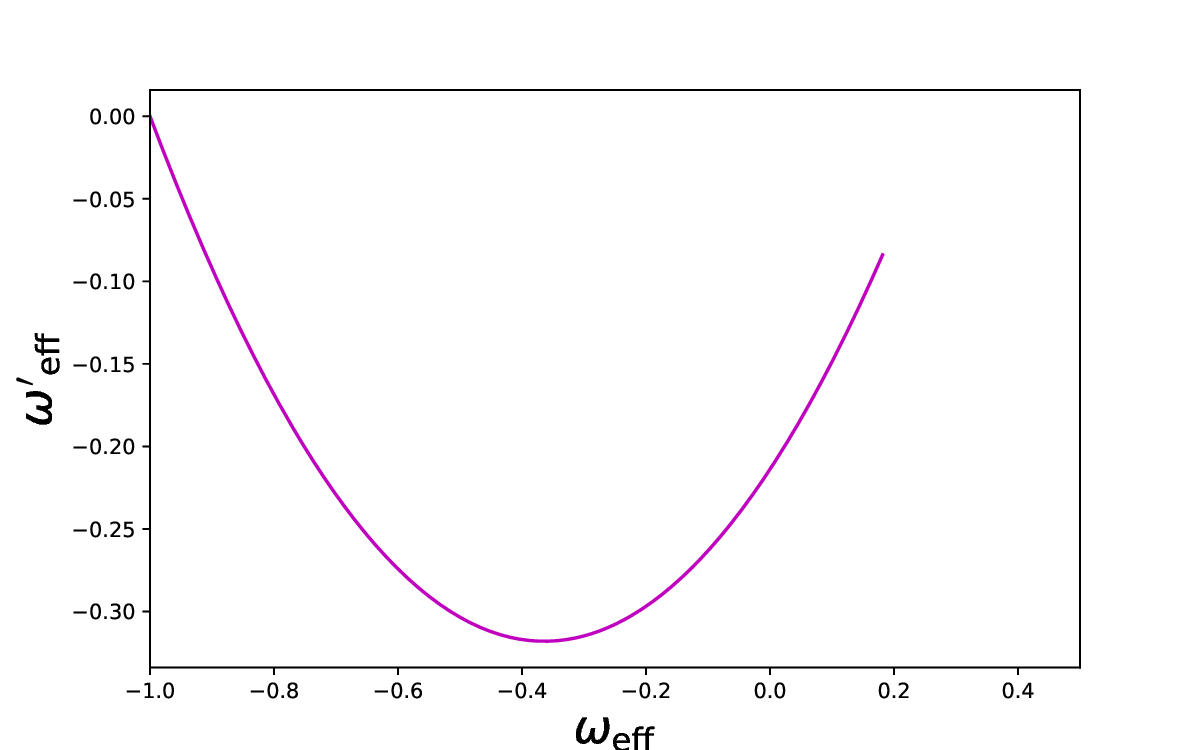}
		\caption{For $n=0.62$}
	\end{subfigure}
	\hfill
	\begin{subfigure}{0.32\textwidth}
		\centering
		\includegraphics[width=\linewidth]{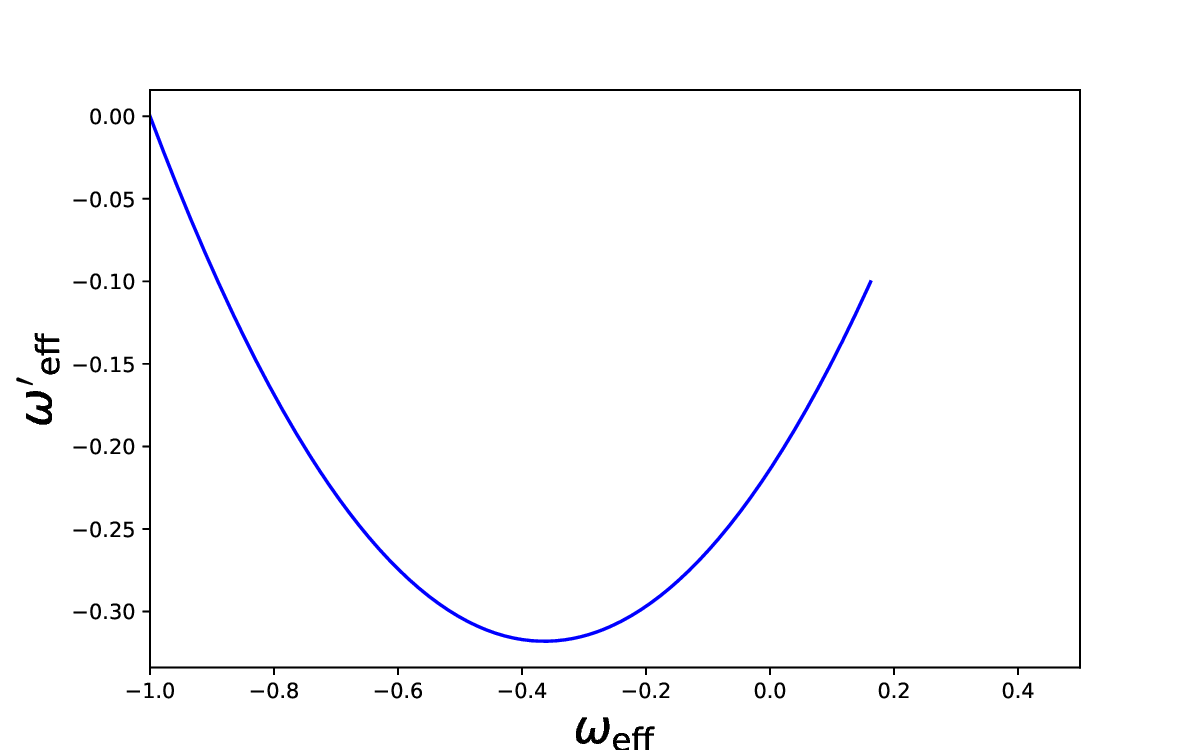}
		\caption{For $n=0.70$}
	\end{subfigure}
	\hfill
	\begin{subfigure}{0.32\textwidth}
		\centering
		\includegraphics[width=\linewidth]{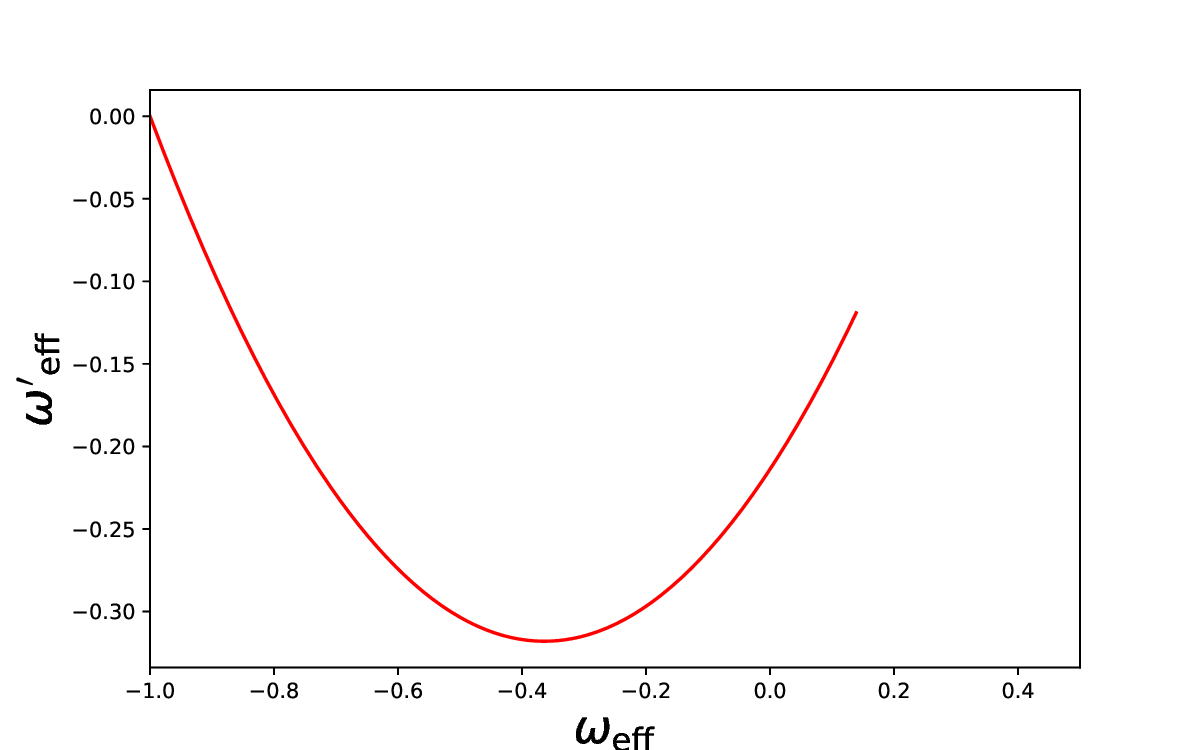}
		\caption{For $n=0.78$}
	\end{subfigure}
	
	\caption{The evolution in the $\omega_{\rm eff}-\omega'_{\rm eff}$ plane for different values of the model parameter $n$.} 
	
	\label{fig:6}
\end{figure}

In this subsection, we have perform a dynamical analysis of $\omega_{\rm eff}$ and its derivative ($\omega'_{\rm eff}$) with respect to $\ln a(t)$ for the $\Lambda_{\mathrm{PVL}}$ model.
This method is considered as a widely recognized tool for investigating dark energy models. 
The $\omega_{\rm eff}$ versus $\omega'_{\rm eff}$ plane effectively highlights the regions corresponding to the accelerated expansion of the universe. This approach was first introduced by Caldwell and Linder~\cite{caldwell2005limits} to study quintessence scalar fields. It provides a robust framework for classifying dark energy models based on their parameter evolution. Two characteristic behaviors are distinguished on this plane: the thawing region ($\omega'_{\rm eff} > 0$ and $\omega_{\rm eff} < 0$) and the freezing region ($\omega'_{\rm eff} < 0$ and $\omega_{\rm eff} < 0$).
\vspace{0.2cm}\\
We computed $\omega'_{\rm eff}$ for the present model and illustrated its evolution in Fig.~(\ref{fig:6}) by differentiating equation~(\ref{35}) with respect to $\ln a(t)$. The dynamical evolution of the model, visualized in the $\omega_{\rm eff}-\omega'_{\rm eff}$ plane in figure~(\ref{fig:6}), shows trajectories for parameter values $n=0.62, 0.70$ and $0.78$. In each case, the paths converge to the same fixed point: $(\omega_{\rm eff},\omega'_{\rm eff})=(-1,0)$. This indicates a stable de Sitter attractor at late times, equivalent to the cosmological constant in the $\Lambda$CDM limit. Throughout their evolution all trajectories remain in the freezing region, confirming a dark energy behavior that aligns with observed cosmic acceleration. Therefore, our model agrees well with facts and shows cosmic acceleration in the freezing zone.
\subsection{Thermodynamic analysis}\label{sec:5.2}
Thermodynamics has a profound and fundamental connection with gravitational theories governing the dynamics of the universe. Jacobson~\cite{jacobson1995thermodynamics} showed that the Einstein field equations can be derived from the laws of black hole thermodynamics, while Padmanabhan~\cite{padmanabhan2002classical} demonstrated that the first law of thermodynamics naturally emerges from the Einstein field equations. These pioneering studies establish a deep correspondence between gravitation and thermodynamics, providing  a robust theoretical framework for thermodynamic investigations in gravitational theories.
The generalized second law of thermodynamics (GSLT) constitutes a cornerstone of such analyses and has been extensively examined within Einstein gravity as well as in various modified theories of gravity~\cite{momeni2016generalized,mamon2021dynamics,pinki2023new,karami2011generalized}. According to the GSLT, the total entropy of the universe, defined as the sum of the horizon entropy and the entropy of the matter content enclosed within the horizon, must be a non-decreasing function of cosmic time. The validity of the GSLT has been explored in both pure BD theory~\cite{duary2020brans} and its modified extensions~\cite{bhattacharya2011brans}.
In this work, we consider dynamical vacuum energy models characterized by a time-varying cosmological constant within the framework of BD theory. It is therefore of particular interest to examine the validity of the GSLT for the $\Lambda_{\mathrm{PVL}}$ model in the context of BD theory. We now analyze the thermodynamic behaviour of the present model through the GSLT. The total entropy of the universe is given by
\begin{equation} {\label{43}}
	S_{tot} = S_{h} + S_{in},
\end{equation}
where, $S_{tot}$ denotes total entropy, $S_{h}$ represents the horizon entropy and $S_{in}$ corresponds to the entropy of total fluid inside the horizon.
\vspace{0.2cm}\\ 
In order to test the validity of the GSLT for our model, we compute the time derivative of the total entropy $S_{tot}$. If the model satisfies the GSLT then we must observe $\dot{S}_{tot}$ $\geq$ 0. We obtain time variation of total entropy $S_{tot}$ form Eq.~(\ref{43}) as
\begin{equation} {\label{44}}
	\dot{S}_{tot} = \dot{S}_{h} + \dot{S}_{in},
\end{equation}
where over-dot represents the time derivative. We now require explicit expressions for the horizon entropy $S_{h}$ and the entropy of the fluid inside the horizon $S_{in}$.
In that study, the entropy associated with the dynamical apparent horizon was considered rather than that of the teleological event horizon. For a dynamical apparent horizon, the horizon entropy is given by $S_{h}=2\pi A$, where $A= 4\pi R_{h}^{2}$ denotes the area of the apparent horizon. Here, $R_{h}$ represents the radius of the apparent horizon, which for a spatially flat FLRW universe is given by $R_{h}= \frac{1}{H}$. Consequently, the entropy of the apparent horizon takes the form
\begin{equation} {\label{45}}
	S_{h} = \frac{8\pi^{2}}{H^{2}}
\end{equation}
and it’s rate of change is given by
\begin{equation} {\label{46}}
	\dot{S}_{h} = -16\pi^{2}\frac{\dot{H}}{H^{3}}.
\end{equation}
The Gibbs law of thermodynamics for fluid inside the horizon produces the relation
\begin{equation} {\label{47}}
	T_{in} dS_{in} = dE_{in}+ p_{t} dV_{h},
\end{equation}
where the subscript $t$ represents the total quantity and $V_{h}=\frac{4}{3} \pi R_{h}^{3}$ is the volume enclosed by the apparent horizon. Now, the rate of change in entropy of the fluid inside the horizon can be obtained as
\begin{equation} {\label{48}}
	\dot{S}_{in} = \frac{(\rho_{t}+ p_{t})\dot{V}_{h}+ \dot{\rho}_{t} V_{h}}{T_{in}}.
\end{equation}
Assuming thermal equilibrium between the cosmic fluid and the horizon, the temperatures of the fluid inside the horizon ($T_{in}$) and the dynamical apparent horizon ($T_{h}$) are same~\cite{duary2020brans}. This temperature corresponds to the Hayward-Kodama temperature and is given by
\begin{equation} {\label{49}}
     T_{h} = \frac{2H^{2}+\dot{H}}{4\pi H}.
\end{equation}
It can be observed that this temperature reduces to the Hawking temperature, $T_{Hawking}= \frac{H}{2\pi}$~\cite{hawking1974black} in de-Sitter spacetime where $\dot{H}=0$. Now, the rate of change of entropy of fluid inside the horizon can be obtained using equations~(\ref{48}) and~(\ref{49}) as
\begin{equation} {\label{50}}
	\dot{S}_{in} = 16\pi^{2}\frac{\dot{H}}{H^{3}}\left(1+\frac{\dot{H}}{2H^{2}+\dot{H}}\right).
\end{equation}
Employing equations~(\ref{44}), ~(\ref{46}) and ~(\ref{50}), we obtain the rate of change of the total entropy as
\begin{equation} {\label{51}}
	\dot{S}_{tot} = \frac{\left(\frac{4\pi \dot{H}}{H^{2}}\right)^{2}}{H\left(\frac{\dot{H}}{H^{2}}+2\right)}.
\end{equation}
\begin{center}
	\begin{figure}
		\includegraphics[width=15.5cm, height=7cm]{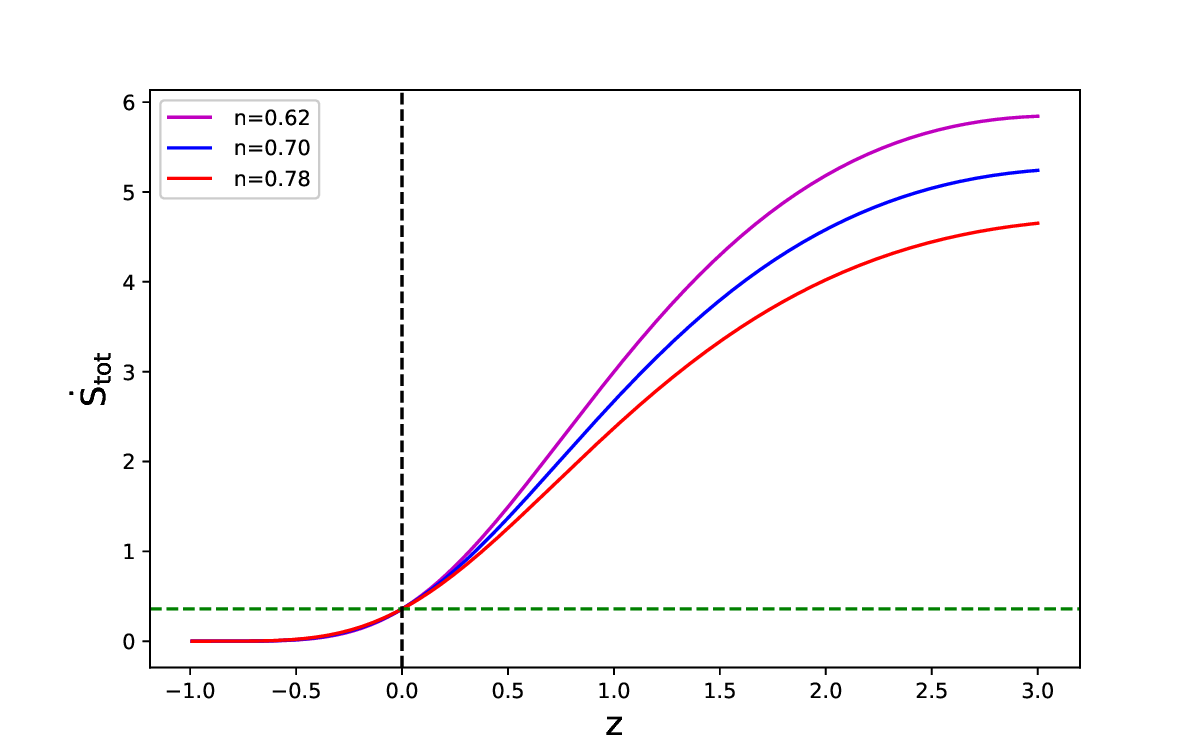}
		\caption{$\dot{S}_{tot}$ versus $z$.} 
		\label{fig:7}
	\end{figure}
\end{center}
If the GSLT is satisfied in a given cosmological model, the time derivative of the total entropy must obey the condition $\dot{S}_{tot}$ $\geq$ 0; otherwise, the model violates the GSLT. For the $\Lambda_{\mathrm{PVL}}$ model, we find that the inequality $\dot{S}_{tot}$ $\geq 0$ is fulfilled provided that $\frac{\dot{H}}{H^{2}}+2$ $\geq 0$. Since the Hubble parameter $H$ remains positive throughout the present and future epochs, consistent with the accelerated expansion of the universe, this condition reduces to $\frac{\dot{H}}{H^{2}}$ $\geq -2$. As the present model is constructed to describe the late-time evolution of the universe, our analysis is restricted to the present epoch and future cosmic evolution. The evolution of $\dot{S}_{tot}$ (given in Eq.~(\ref{51})) as a function of redshift $z$ is shown in Fig.~(\ref{fig:7}). From the figure~(\ref{fig:7}), it is evident that the GSLT remains valid during the present epoch and into the future for three values of the parameter $n$. Further, we observe that $\dot{S}_{tot}$ freezing to zero in far future, i.e. $\dot{S}_{tot} \to 0$ as $z \to -1$. It is interesting to observe that $S_{tot}$ becomes almost constant in far future.
\section{Conclusions}\label{sec:6}
In this work, we have investigated the late-time cosmological evolution of a spatially flat, homogeneous and isotropic FLRW universe within the framework of BD theory by incorporating a time-dependent vacuum energy density. The matter sector is described by a pressureless perfect fluid, while the BD scalar field is assumed to evolve as a power law of the scale factor, $\phi \varpropto a^{\epsilon}$, ensuring compatibility with observational constraints on the variation of the gravitational coupling. Two different vacuum decay scenarios were considered: the hybrid vacuum law $\Lambda(t)=\alpha H^{2}+\beta \dot{H}$ and the power vacuum law $\Lambda(H)=\alpha_{1}H^{n}$, where $\alpha$, $\beta$, $\alpha_{1}$ and $n$ are model parameters in order to assess their ability to describe the observed cosmic expansion.
\vspace{0.2cm}\\ 
The $\Lambda_{\mathrm{HVL}}$ model leads to a power-law expansion of the scale factor characterized by a constant deceleration parameter. As a consequence, this model is unable to account for the observed transition from an early decelerated expansion phase to the present accelerated epoch and is therefore disfavored from a phenomenological perspective. In contrast, the $\Lambda_{\mathrm{PVL}}$ model naturally produces a redshift dependent deceleration parameter, allowing for a smooth transition from a matter-dominated decelerated era to late-time accelerated expansion.
\vspace{0.2cm}\\ 
For the $\Lambda_{\mathrm{PVL}}$ model, the values of the transition redshift $z_{tr}$ and the present-day deceleration parameter $q_{0}$ for three representative values of the parameter $n$ are summarized in Table~\ref{table:1},
providing clear evidence for late-time cosmic acceleration. At the present epoch ($z=0$), the effective EoS parameter exhibits a quintessence-like dark energy behavior with its present-day values $\omega_{0}$ listed in Table~\ref{table:1}
and asymptotically approaches the $\Lambda$CDM limit, $\omega_{\mathrm{eff}}\rightarrow -1$ in the far future corresponding to a de Sitter-like phase of expansion.
\vspace{0.2cm}\\ 
In the asymptotic future $(z \to -1)$, the $\Lambda$CDM scenario predicts the cosmological parameters $q=-1$, $\omega=-1$, $j=1$ and snap (s)$=1$, which collectively signify a de Sitter-like phase of cosmic expansion. An equivalent outcome is obtained in the de Sitter model, where the Hubble parameter remains constant, yielding identical values of $q=-1$, $j=1$ and $s=1$. These results reflect a stable late-time expansion and underscore the consistency of de Sitter dynamics as the asymptotic state of the universe.
Ultimately, cosmographic analysis based on the jerk and snap parameters further supports the viability of the model and the present-day values of the jerk $(\mathit{j}_{0})$ and snap $(\mathit{s}_{0})$ parameters are provided in Table~\ref{table:1},
indicate a mild deviation from the standard $\Lambda$CDM scenario at the current epoch while the late-time evolution shows convergence toward the $\Lambda$CDM behaviour, reinforcing the consistency of the model with the expected asymptotic cosmic dynamics. In addition, the evolution of the Om($\mathit{z}$) diagnostic for the $\Lambda_{\mathrm{PVL}}$ model with three values of the parameter $n$ favors a quintessence-like dark energy behaviour at the present epoch. According to the proposed model, the present cosmic age $t_{0}$ is provided in Table~\ref{table:1}. The trajectories in $\omega_{\rm eff}-\omega'_{\rm eff}$ plane shows that the model evolves in the freezing region and approaches the vicinity of the $\Lambda$CDM model in the future, which has values $\omega_{\rm eff}=-1$, $\omega'_{\rm eff}=0$. We performed a thermodynamic analysis to test the validity of the GSLT within the $\Lambda_{\mathrm{PVL}}$ model. By examining the evolution of the total entropy rate $\dot{S}_{tot}$ as a function of redshift, we found that the GSLT remains satisfied during the present epoch as well as in the future for three values of the parameter $n$. Thus, the model is thermodynamically viable and successfully passes the GSLT criterion.
\vspace{0.2cm}\\ 
In conclusion, the power vacuum law model within BD theory provides a viable and phenomenologically consistent description of the universe’s expansion history. The combined effects of a dynamical vacuum energy density and a varying gravitational coupling offer a compelling alternative framework for explaining the observed late-time acceleration of the universe.
\section*{\textbf{Acknowledgements}}
One of the authers, G. P. Singh gratefully acknowledges the support provided by the Inter-University Centre for Astronomy and Astrophysics (IUCAA), Pune, India, under the Visiting Associateship Programme.


\end{document}